\documentclass[a4paper,10pt]{article}
\usepackage{graphicx}
\usepackage{comment}
\usepackage{amssymb}
\usepackage{amsmath}
\usepackage{nicefrac}
\usepackage{graphicx}
\usepackage{dcolumn}
\usepackage{bm}
\usepackage{comment}
\usepackage{multirow}
\usepackage{cite}
\usepackage{mathrsfs}
\usepackage{subcaption}
\usepackage{xcolor}

\begin{document}

\huge

\begin{center}
Issues in the calculations of dc conductivity of warm dense aluminum
\end{center}

\vspace{0.5cm}

\large

\begin{center}
Nadine Wetta$^{a,}$\footnote{nadine.wetta@cea.fr} and Jean-Christophe Pain$^{a,b}$
\end{center}

\normalsize

\begin{center}
\it $^a$CEA, DAM, DIF, F-91297 Arpajon, France\\
\it $^b$Universit\'e Paris-Saclay, CEA, Laboratoire Mati\`ere en Conditions Extr\^emes,\\
\it 91680 Bruy\`eres-le-Ch\^atel, France
\end{center}

\abstract{In the Ziman-Evans formulation, the electrical resistivity involves  several quantities like the plasma mean ionization and chemical potential, the electron-ion scattering cross-section, the ion-ion structure factor, and the derivative of the Fermi distribution with respect to electron energy. Therefore, in order to make significant progress while comparing different models or analyzing experimental measurements, it is important to try to get insight into such partial physical quantities. In the present work, we propose to investigate the sensitivity of the resistivity to the different physical quantities required for its computation.}

\footnotetext{\textbf{Abbreviations:} AA, Average-Atom; dc, direct current; fcc, face-centered cubic;  GGA, Generalized Gradient Approximation; HNC, HyperNetted-Chain; HSE, Heyd–Scuseria–Ernzerhof; KS, Kohn-Sham; KSDT, Karasiev-Sjostrom-Dufty-Trickey; LCLS, Linac Coherent Light Source; LDA, Local-Density Approximation; NPA, Neutral Pseudo-Atom; MC, Monte-Carlo; OCP, One-Component Plasma; PBE, Perdew-Burke-Ernzerhof; PDW, Perrot-Dharma-wardana; PZ, Perdew-Zunger; QMD, Quantum Molecular Dynamics;  WDM, Warm Dense Matter; xc, exchange-correlation.}

\section{Introduction}\label{sec1}

Recently, we presented an approach allowing for a consistent calculation of electrical conductivity of dense matter from the solid state to the hot plasma. The formalism relies on a unified procedure, consisting in dropping elastic scattering contributions to solid's and liquid's structure factors in the framework of the Ziman theory \cite{Wetta2020}. The solid's structure factor was computed using a multi-phonon expansion. For the liquid, a similar elastic contribution, signature of a long-range order persisting during the characteristic electron-ion scattering time, was removed from the structure factor. All the quantities required for the calculation of the resistivities are obtained from our Average-Atom (AA) model {\sc Paradisio}, including the total HyperNetted-Chain (HNC) structure factor used from the liquid state to the plasma. In that way, no interpolation between two limiting structure factors is required. We derived the correction to apply to the resistivity in order to account for the transient long-range order in the liquid and found that it improves considerably the agreement with Quantum Molecular Dynamics (QMD) simulations and experimental aluminum's isochoric and isobaric conductivities.

The resistivity itself involves several ingredients, such as the ion-ion structure factor, the plasma mean ionization and chemical potential, the electron-ion scattering cross-section and the derivative of the Fermi distribution with respect to electron energy. Therefore, comparing different models or trying to understand potential discrepancies between experiment and theory is difficult. In the present work, we propose to investigate the sensitivity of the resistivity to the different physical quantities required for its computation in the framework of Ziman's theory.

In section \ref{sec2}, the main features of the Ziman-Evans formulation of electrical resitivity are recalled, together with the above mentioned correction to the resistivity. The effect of the definition of the mean ionization on direct-current (dc) conductivity is investigated in section \ref{sec3}. The impacts of the choice of the finite-temperature exchange-correlation (xc) functional and of the ionic structure factor are studied in sections \ref{sec4} and \ref{sec5} respectively. The sensitivity of the interpretation of two-temperature experiments with respect to the ionic structure factor is discussed in section \ref{sec6}.

\section{Ziman formula and the Average-Atom model}\label{sec2}

Although this is not necessary under the thermodynamic conditions considered in this work, the following formulas will be given in the relativistic formalism, for the sake of consistency with the relativistic AA code {\sc Paradisio} \cite{Penicaud2009} that was used to provide the needed inputs for resistivity calculations. All formulas will be given in atomic units (i.e. $e=\hbar=m_e=1$).\\
\indent The Ziman formulation of the electrical resistivity \cite{Ziman1961} describes, within the linear response theory, the acceleration of free electrons in a metal and their scattering by an ion. The resistivity reads then
\begin{equation}\label{eta}
    \eta=-\dfrac{1}{3\pi {Z^*}^2 n_i} \int_0^\infty \dfrac{\partial f}{\partial \epsilon}(\epsilon,\mu^*) I(\epsilon) d\epsilon,
\end{equation}
where $n_i$ is the ion density, and $Z^*$ the mean ionic charge. The Fermi-Dirac distribution and its derivative read respectively
\begin{equation}\label{fermidirac}
    f(\epsilon,\mu^*)=\dfrac{1}{e^{\beta (\epsilon-\mu^*)}+1}
\end{equation}
and
\begin{equation}
    \dfrac{\partial f}{\partial\epsilon}(\epsilon,\mu^*)=-\beta f(\epsilon,\mu^*)\left[1-f(\epsilon,\mu^*)\right], 
\end{equation}
where $\beta=1/(k_B T)$ and $\mu^*$ denotes the chemical potential associated to the free electron gas of density $n_e=Z^* n_i$, given by
\begin{equation}
    \dfrac{2}{(2\pi)^3} \int_0^\infty f(\epsilon,\mu^*)\,4\pi k^2 dk=Z^*,
\end{equation}
or
\begin{equation}
    \mathscr{F}_{1/2}(\beta\mu^*)=\dfrac{\pi^2}{\sqrt{2}}\beta^{3/2}Z^*,
\end{equation}
introducing the Fermi function of order $1/2$ 
\begin{equation}
    \mathscr{F}_{1/2}(x)=\int_0^\infty \dfrac{t^{1/2}}{(1+e^{t-x})}dt.
\end{equation}
The function $I(\epsilon)$ is given by
\begin{equation}
    I(\epsilon)=\int_0^{2k}q^3 S(q) \Sigma(q) dq,
\end{equation}
where $S(q)$ denotes the static ion-ion structure factor and $\Sigma(q)$ the scattering cross-section. The vector $\vec{q}=\vec{k}^\prime-\vec{k}$ is the momentum transferred in the elastic scattering event (i.e. such as $|\vec{k}^\prime|=|\vec{k}|$) of a conduction electron from an initial state $\vec{k}$ to a final $\vec{k}^\prime$ one. Introducing the scattering angle $\theta\equiv (\vec{k},\vec{k}^\prime$) and its cosine $\chi=\cos\theta$, one has $q^2=2k^2 (1-\chi)$ and
\begin{equation}
    I(\epsilon)=2k^4 \int_{-1}^1 S\left[k\sqrt{2(1-\chi)}\right]|a(k,\chi)|^2 (1-\chi) d\chi.
\end{equation}
Energy $\epsilon$ and momentum $k$ are related (within the relativistic formalism, $c$ being the speed of light, and using atomic units) by
\begin{equation}
    k=\sqrt{2\epsilon \left(1+\dfrac{\epsilon}{2c^2}\right)}.
\end{equation}
The $t-$matrix formalism of Evans \cite{Evans1973} provides the electron-ion scattering amplitude $|a(k,\chi)|$, whose square is actually $\Sigma(q)$, given by, in the relativistic framework \cite{Sterne2007}
\begin{equation}\label{scattering}
    |a(k,\chi)|^2=\frac{1}{k^2}\left(\Big|\sum |\kappa|e^{i\delta_\kappa(k)}\sin[\delta_\kappa(k)]P_{\ell}(\chi)\Big|^2+\Big|\sum \frac{|\kappa|}{i\kappa}e^{i\delta_\kappa(k)}\sin[\delta_\kappa(k)]P^1_\ell(\chi)\Big|^2\right),
\end{equation}
where the summations are performed over the electronic states, labelled by the relativistic quantum number $\kappa$, which is related to the quantum number $\ell$ associated to the orbital momentum $L$ and to the spin $s$ by the relations
\begin{equation}
    \begin{array}{l c l}
        \kappa=-(\ell+1)& \mathrm{ for }& s=+1/2,\\
        \kappa=\ell& \mathrm{ for }& s=-1/2.
    \end{array}
\end{equation}
The functions $P_\ell$ and $P^1_\ell$ denote respectively the Legendre and associated Legendre polynomials.

The average ion charge $Z^*$ and phase-shifts $\delta_{\kappa}(k)$ needed in respectively Equations (\ref{eta}), (\ref{fermidirac}) and (\ref{scattering}), can be obtained with the help of AA codes. The ionic structure factor $S(k)$ is usually obtained independently.

\begin{figure}
    \centering
    \includegraphics[width=0.65\textwidth]{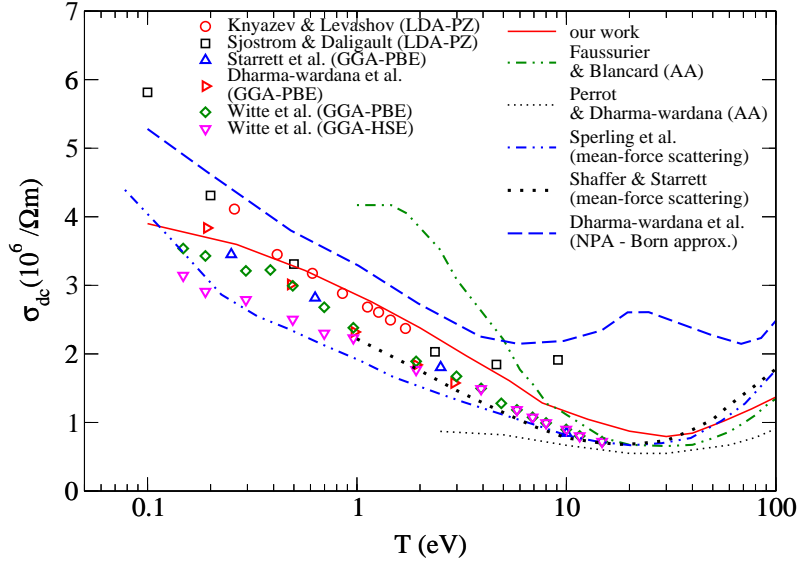}
    \caption{Static electrical conductivity of solid-density aluminum as a function of temperature. Symbols: Quantum Molecular Dynamics calculations \cite{Knyazev2013,Sjostrom2015,Starrett2020,Witte2018,Dharmawardana2017} using various exchange-correlation functionals; within Local-Density Approximation (LDA): Perdew-Zunger (PZ)\cite{Perdew1981}, in the framework of the Generalized Gradient Approximation (GGA):  Perdew-Burke-Ernzerhof (PBE)\cite{PBE1996} and  Heyd–Scuseria–Ernzerhof (HSE)\cite{HSE2003}. Red line: our work \cite{Wetta2020}. Other lines: other Average-Atom results: Perrot and Dharma-wardana \cite{Perrot1999}, Faussurier and Blancard \cite{Faussurier2019}, approaches based on the mean-force scattering concept (Shaffer and Starrett \cite{Shaffer2020}, Sperling {\it et al.}\cite{Sperling2017}), and Neutral Pseudo-Atom (NPA) model of Dharma-wardana {\it et al.}\cite{Dharmawardana2017}.}
    \label{figcondAl1}
\end{figure}

Figure \ref{figcondAl1} presents a set of theoretical electrical conductivities for aluminum at solid density and temperatures ranging from 0.1 eV up to 100 eV. Symbols correspond to QMD calculations using different xc functionals, and the lines to some AA  ones. Our own results obtained with our AA code {\sc Paradisio} \cite{Penicaud2009} are given by the red curve. {\sc Paradisio} is based on Liberman's atom-in-jellium model {\sc Inferno} \cite{Liberman1979} and handles all electronic states (bound and continuum ones) on an equal footing within the framework of quantum mechanics, which is an essential condition to correctly incorporate the effects of electron-electron interactions in the electron-ion phase-shifts \cite{Pain2007}. We recently applied it to the study of aluminum's thermal electronic properties \cite{Wetta2019} and showed its efficiency at low temperatures, where the atom-in-jellium approximation is often considered as too crude. In our work, we retained the number of charges in the continuum states for the ion charge $Z^*$, applied the HNC closure relation to the system of charged spheres for the ionic structure factor $S(k)$, and used KSDT (Karasiev-Sjostrom-Dufty-Trickey) finite-temperature Local-Density Approximation (LDA) exchange-correlation functional \cite{Karasiev2014}. As we will show later, using the KSDT functional, our results are close to QMD simulations based on functionals at the same LDA level. We also assumed the persistence of long-range order in the liquid and therefore applied the following correction to Ziman's resistivity \cite{Wetta2020} 
\begin{equation}
    \delta\eta=-\dfrac{1}{3\pi {Z^*}^2 n_i}\sum_G \dfrac{N(G)}{4\pi}\mathrm{e}^{-2W(G)}
    \times \int_{G/2}^\infty \left(-\dfrac{\partial f}{\partial k} \right) k^2 G^2 \times\Bigg|a\left(k,1-\dfrac{G^2}{2k^2}\right)\Bigg|^2 dk,
\end{equation}
where $N(G)$ denotes the number of reciprocal lattice vectors of same length $G$ and $\mathrm{e}^{-2W(G)}$ the Debye-Waller factors accounting for thermal decay of the long-range order. This correction results from the extension to liquids of a prescription by Rosenfeld and Stott initially formulated for solids \cite{Rosenfeld1990}, which consists in removing the contribution of the perfect rigid lattice from the total structure factor used in Ziman's formula. This concept has been precedently applied by Baiko {\it et al.} in the framework of astrophysics \cite{Baiko1998}. An experimental proof of transient long-range order in melted gold has been presented \cite{Mo2018} through the coexistence of Debye-Scherrer rings and Laue diffraction peaks in x-ray diffraction patterns at times exceeding the electron-ion equilibration one. In our precedent work on aluminum, the correction $\delta\eta$ resulting from this extension was essential to explain experimental aluminum isobaric electrical conductivities.

The other curves in Figure \ref{figcondAl1} reproduce the results of other studies, among which AA calculations with different choices for $Z^*$, xc functional or ionic structure factor $S(k)$ \cite{Perrot1999,Faussurier2019}. For information, we also reported other type of calculations like those based on mean-force scattering models \cite{Sperling2017,Shaffer2020}, giving results comparable to AA ones. The Neutral Pseudo-Atom (NPA) calculation of Dharma-wardana {\it et al.} \cite{Dharmawardana2017}, represented by the blue dashed curve, differs by the use of other boundary conditions. Indeed, while most AA models (including {\sc Paradisio}) place the boundary at the surface of the Wigner-Seitz sphere, NPA extends it to a correlation sphere, on the surface of which electron-electron as well as ion-ion correlations are cancelled out. Dharma-wardana {\it et al.}'s calculations also differ by the use of pseudo-potentials and of the Born approximation in the calculation of the scattering cross-section.

\begin{figure}
    \centering
    \includegraphics[width=0.65\textwidth]{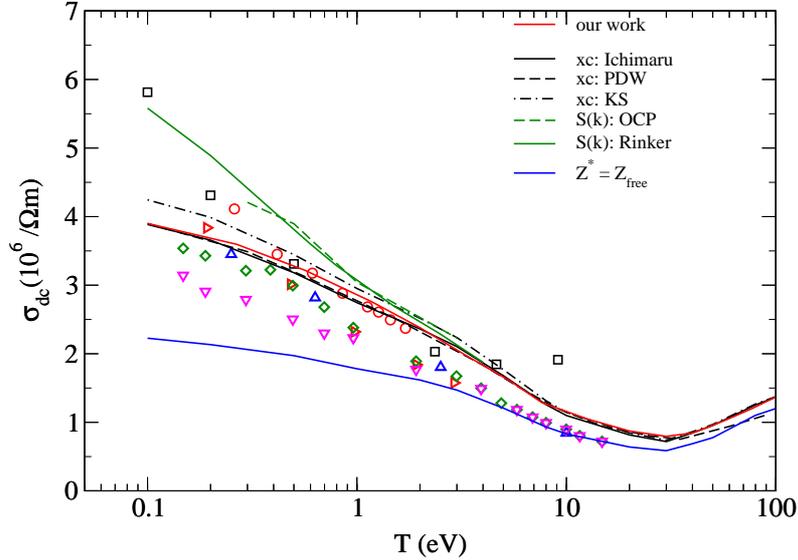}
    \caption{The impact of the exchange-correlation functional on aluminum's dc conductivity at solid density $\rho_0=2.7$ g/cm$^3$ is illustrated by the black curves (PDW: Perrot-Dharma-wardana\cite{Perrot2000,Perrot_erratum}, KS: Kohn-Sham\cite{Kohn1965}), that of the ion charge $Z^*$ by the blue line, and the impact of the ionic structure factor $S(k)$ (OCP: One-Component Plasma, Rinker: Equation (\ref{rinker})) by the green ones. The red curve is our result for $Z^*=Z_\mathrm{cont}$, Karasiev {\it et al.}'s exchange-correlation functional \cite{Karasiev2014}, and HyperNetted-Chain ionic structure factor. The legend for the symbols is the same as in Figure \ref{figcondAl1}.}
    \label{figcondAl2}
\end{figure}

We present in Figure \ref{figcondAl2} some calculations using {\sc Paradisio} with some alternative choices for $Z^*$, $S(k)$ and xc functional. We will analyze further how these choices impact Ziman's resistivity in the next two sections.

\section{Effect of mean ionization on DC conductivity within Ziman's approach}\label{sec3}

\begin{figure}[t]
    \centerline{
    \includegraphics[width=0.65\textwidth]{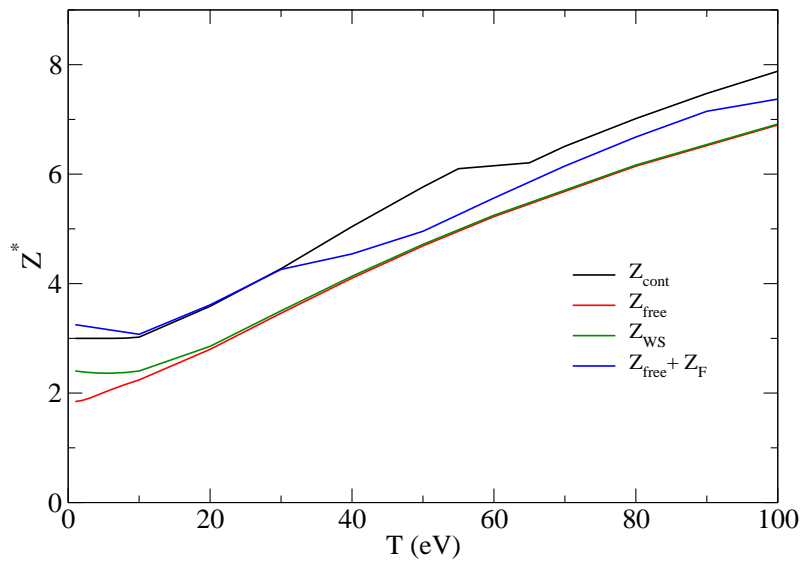}}
    \caption{Aluminum at solid density: some possible definitions of the ion charge $Z^*$ within ion-sphere models. $Z_\mathrm{cont}$ is the number of continuum electrons (in black), $Z_\mathrm{free}$ denotes the ideally free electrons alone (in red), $Z_\mathrm{WS}$ the charge on the Wigner-Seitz sphere (in green). Finally, the blue curve represents the sum $Z_\mathrm{free}+Z_F$, where $Z_F$ is the contribution of the Friedel oscillations outside the ion sphere.
\label{figZstar}}
\end{figure}
\begin{figure}[t]
    \centerline{\includegraphics[width=0.65\textwidth]{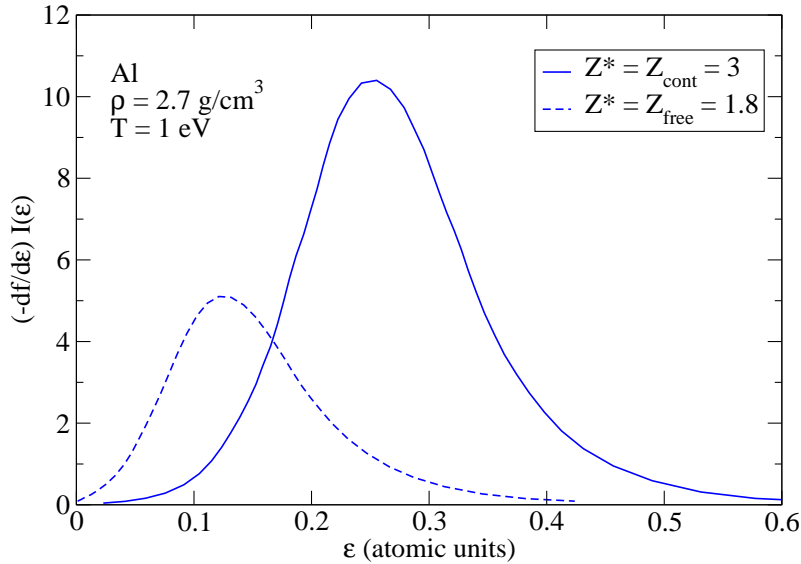}}
    \caption{Solid density aluminum at $T=1$ eV: impact of the ion charge $Z^*$ on the integrand $\left( -\frac{\partial f}{\partial\epsilon}\right)I(\epsilon)$. Full line: $Z^*=Z_\mathrm{cont}$, and dashed line: $Z^*=Z_\mathrm{free}$. Both reduction of $Z^*$ and the associated shift of the chemical potential partially counter the augmentation of the factor $1/Z^{*2}$ in Ziman's formula for the resistivity. \label{figZ*}}
\end{figure}
\indent As shown in Figure \ref{figcondAl2}, the ion charge $Z^*$ is one of the quantities that impacts the most Ziman's electrical resistivity. Since $Z^*$ is not the average value of a quantum mechanical operator, it does not have a clear and unique definition. We give here three possible ones among the most widely used \cite{More1985}. First, $Z^*$ can be identified to the total number of continuum electrons 
\begin{equation}\label{Z*}
    Z_\mathrm{cont}=\int_0^\infty f(\epsilon,\mu)X_\mathrm{cont}(\epsilon)d\epsilon,
\end{equation}
where $X_\mathrm{cont}(\epsilon)$ denotes the continuum density of states, which includes not only electrons in the ideal (i.e. non-localized) states but also possible ``quasi-bound'' electrons in resonances. The chemical potential $\mu$ (not to be confused with $\mu^*$) is given by the electrical neutrality condition inside the ion sphere, which reads, in the framework of the {\sc Inferno} model
\begin{equation}
    Z=\int_0^\infty d\epsilon\,f(\epsilon,\mu) \left[X_\mathrm{bound}(\epsilon)+X_\mathrm{cont}(\epsilon)\right],
\end{equation}
where $X_\mathrm{bound}(\epsilon)$ is the density of bound states and $Z$ denotes the atomic number. Alternatively, $Z^*$ may be specified as the ideally free charges in the continuum only
\begin{equation}
    Z_\mathrm{free}=\int_0^\infty f(\epsilon,\mu)X_\mathrm{ideal}(\epsilon) d\epsilon,
\end{equation}
where the ideal density of state reads, within the relativistic framework
\begin{equation}
    X_\mathrm{ideal}(\epsilon)=\dfrac{k(1+\epsilon/c^2)}{\pi^2 n_i}.
\end{equation}

Actually, the ideal electron density is the one for $r\rightarrow\infty$, and so
\begin{equation}
    Z_\mathrm{free}=\left(\lim_{r\rightarrow\infty}n_e(r)\right)/n_i.
\end{equation}
Within the framework of the {\sc Inferno} model, the density of charge $n_e(r)$ is constant at $r>R_\mathrm{WS}$, where $R_\mathrm{WS}$ denotes the radius of the Wigner-Seitz sphere
\begin{equation}
    R_\mathrm{WS}=\left(\frac{3}{4\pi n_i}\right)^{1/3},
\end{equation}
so that $Z_\mathrm{free}=\overline{\rho}/n_i$, $\overline{\rho}$ being the jellium density.

A third definition of $Z^*$ involves the value of the electron density at the radius $R_\mathrm{WS}$ 
\begin{equation}
    Z_\mathrm{WS}=n_{e}(R_\mathrm{WS})/n_i.
\end{equation}

A fourth definition for $Z^*$ writes
\begin{equation}
    Z^*=Z_\mathrm{free}+Z_F,
\end{equation}
where $Z_F$ denotes the charge displaced by the electron-ion potential, whose value can be obtained applying the finite-temperature Friedel sum rule. In the framework of the relativistic formalism \cite{Friedel1952,Faussurier2021}
\begin{equation}
    Z_F=\dfrac{2}{\pi}\int_0^\infty d\epsilon \left(-\dfrac{\partial f}{\partial\epsilon}\right)\sum_\kappa |\kappa|\delta_\kappa (k).
\end{equation}

The equality of these four values is not obvious for AA models. However, in most situations, they are not so different, and depart significantly from each other only in conditions favorable to pressure ionization, i.e. at high densities, and especially at low temperatures, when resonance states may notably be populated. Within the {\sc Inferno} model, this situation results in a sudden jump of the difference ($Z_\mathrm{WS}-Z_\mathrm{free}$). Since $Z_\mathrm{free}$ and $Z_\mathrm{WS}$ are close to each other for 300 K $\leq T\leq$ 100 eV (see Figure \ref{figZstar}), we conclude that no ``quasi-bound'' states form, and therefore retained $Z^*=Z_\mathrm{cont}$ (Equation (\ref{Z*})) as the definition of the mean ion charge, considering that the supplementary charge (typically one electron in the considered temperature range) contributes to electrical conductivity, although it is not ideally free.

The value retained for $Z^*$ will have a direct impact on Ziman's electrical resistivity by means of the $1/(3\pi {Z^*}^2 n_i)$ factor. For instance, at $T=1$ eV, lowering $Z^*$ from $Z_\mathrm{cont}=3$ to $Z_\mathrm{free}=1.8$ will multiply the resistivity by $(3/1.8)^2\approx 2.78$. The same $\Delta Z^*$ reduction has however the opposite effect on the integrand $-\left(\frac{\partial f}{\partial\epsilon}\right)I(\epsilon)$ of Ziman's expression (Equation (\ref{eta})) as illustrated in Figure \ref{figZ*}. Indeed, the charge of the ions impacts the structure factor and induces a decrease of the value of the integrand while the shift of the chemical potential $\mu^*$ also contributes to reduce the integral.

\section{Impact of the finite-temperature exchange-correlation functional}\label{sec4}

\begin{figure}[t]
    \centerline{\includegraphics[width=0.65\textwidth]{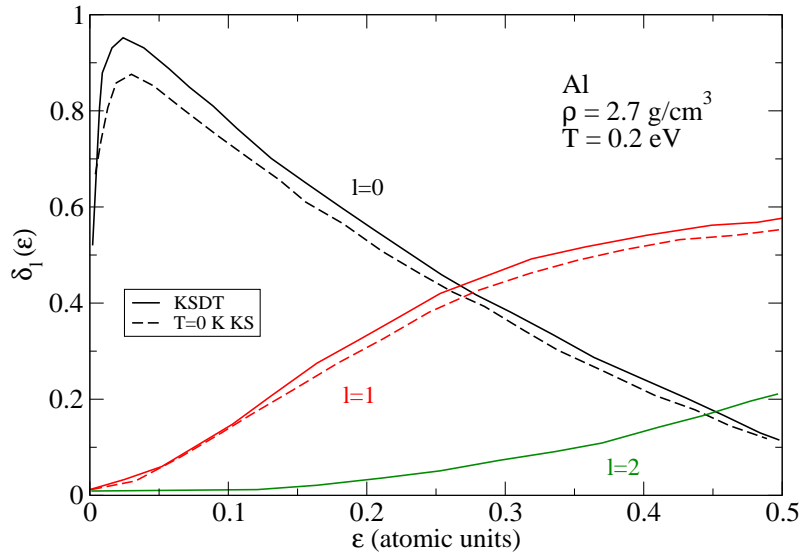}}
    \caption{Solid density aluminum at $T=0.2$ eV: effect of exchange-correlation functional on the $\ell=0, 1$ and 2 phase-shifts $\delta_\ell$. The solid lines correspond to Karasiev {\it et al.}'s finite-temperature (KSDT) functional\cite{Karasiev2014}, and the dashed lines to $T=0$ K Kohn-Sham (KS) exchange functional\cite{Kohn1965}.\label{figxc}}
\end{figure}
\begin{figure}
    \centering
    \includegraphics[width=0.65\textwidth]{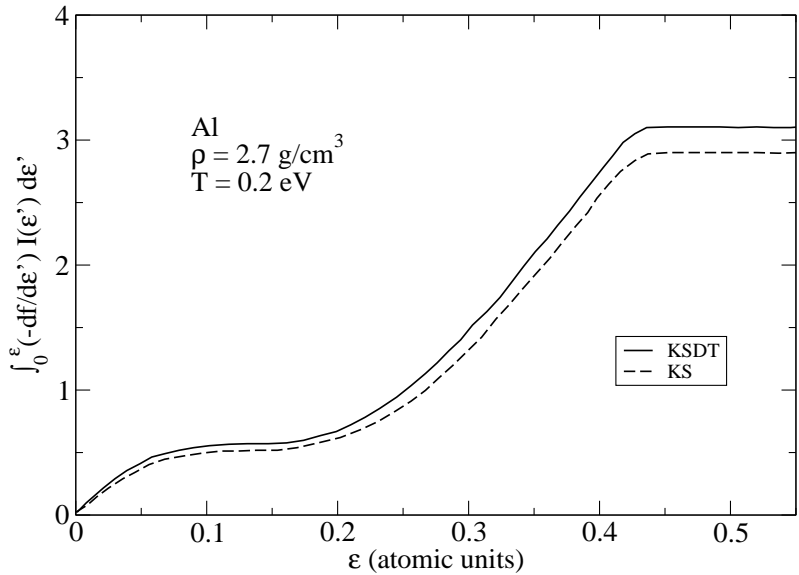}
    \caption{Cumulative integral $\int_0^\epsilon d\epsilon^\prime \left(-\frac{\partial f}{\partial\epsilon^\prime} \right)I(\epsilon^\prime)$. Impact of the finite-temperature electronic exchange-correlation (KSDT:  Karasiev-Sjostrom-Dufty-Trickey\cite{Karasiev2014} and KS: $T=0$ K Kohn-Sham exchange\cite{Kohn1965}).}
    \label{figcumulxc}
\end{figure}
Kohn-Sham (KS) exchange functionals \cite{Kohn1965}, although valid at $T$=0 K, are largely used for the study of hot plasmas. In the framework of this approximation, the exchange potential reads
\begin{equation}
    V_x[n(r)] =-\dfrac{1}{\pi}\left[3\pi^2 n(r) \right]^{1/3}.
\end{equation}
For calculations in the Warm Dense Matter (WDM) conditions, finite-temperature xc functionals should be preferred. {\sc Paradisio} uses Karasiev {\it et al.}'s formulation \cite{Karasiev2014} (designated by KSDT in the literature), built on the same spirit as Ichimaru {\it et al.}'s functional \cite{Ichimaru1987}. Both approaches fit Monte-Carlo (MC) data by Pad\'e approximants, but differ by the quantity which is fitted. Ichimaru {\it et al.} approximated MC interaction energies $E_\mathrm{int}$, and obtained a rather complicated analytical expression for the xc free energy functional $f_\mathrm{xc}$ after integration over the coupling constant
\begin{equation}
    f_\mathrm{xc}=\int_0^\Gamma \dfrac{d\Gamma}{\Gamma}\left(\dfrac{E_\mathrm{int}}{N k_B T} \right)
\end{equation}
of the Pad\'e approximant. The xc potential is then obtained by derivation
\begin{equation}
    V_\mathrm{xc}=\dfrac{\partial}{\partial n} \left(n f_\mathrm{xc}\right),
\end{equation}
or, introducing $r_s=r_\mathrm{WS}/a_B$, $a_B$ being the first Bohr radius %
\begin{equation}
    V_\mathrm{xc}=f_\mathrm{xc}-\dfrac{3}{r_s}\dfrac{\partial f_\mathrm{xc}}{\partial r_s}.
\end{equation}
The KSDT approach fits the discrete $f_\mathrm{xc}$ MC data with the following Pad\'e approximant \cite{Karasiev2014} (where $t=T/T_F$, $T_F$ denoting the Fermi temperature)
\begin{equation}
    f_\mathrm{xc}(r_s,t)=-\dfrac{1}{r_s}\dfrac{a(t)+b(t)~ r_s^{1/2}+c(t)~r_s}{1+d(t)~r_s^{1/2}+e(t)~r_s},
\end{equation}
which is easier to derive analytically than Ichimaru {\it et al.}'s expression. The $a(t), b(t), c(t), d(t)$ and $e(t)$ functions are defined by a set of 18 parameters \cite{Karasiev2014}. Groth {\it et al.} recently revised their values on the basis of new Restricted Path Integral Monte-Carlo  data \cite{Groth2017}.

Perrot and Dharma-wardana proposed a rather different approximant for the xc free energy \cite{Perrot2000,Perrot_erratum} (named PDW). At $T=0$ K, the expression reduces to the sum of the Perdew-Zunger (PZ) correlation energy and an approximant of the exchange energy. The PDW expression is also built to respect two exact leading terms at high temperature, i.e. the Debye correlation term proportional to $T^{-1/2}$, and the exchange high-temperature limit, varying as $T^{-1}$, whereas the KSDT formulation only verifies the first one \cite{Karasiev2014}.

As shown in Figure \ref{figcondAl2}, the three finite-temperature xc functionals lead to quite the same electrical conductivities. At the most, we note some differences for $T\gtrsim 40$ eV between Ichimaru and KSDT (in its original formulation \cite{Karasiev2014} or with the revised parameters \cite{Groth2017}) on one hand, and, on the other hand, the PDW formulation. The fact that PDW imposes that the high-temperature $f_\mathrm{xc}$ approximant has to respect both the Debye correlation free-energy limit $f_c\varpropto T^{-1/2}$ and the exchange free-energy one $f_x\varpropto T^{-1}$ could explain why the PDW curve (black dashes) departs from the Ichimaru and KSDT ones (red and black full lines) at high temperature. Without surprise, the finite-temperature xc functionals mainly impact the conductivities in the WDM regime, as illustrated by the difference between the red and black curves corresponding to finite-temperature functionals and the black dash-dot line representing the ($T=0$ K) KS formulation. Figure \ref{figcondAl2} suggests a reduction of the electrical conductivity as the xc functionals gain in complexity. Indeed, conductivities are the highest with $T=0$ K KS one, smaller with LDA finite-temperature xc functionals (red and black curves: our calculations, red circles and black squares: QMD), and finally the lowest when Generalized Gradient Approximation (GGA) functionals are used (QMD: blue, red, magenta triangles and green losanges).

Figure \ref{figxc} presents the effect of finite-temperature xc on phase-shifts for aluminum at solid density and $T=0.2$ eV (under these conditions $\delta_\mathrm{\kappa=\ell}=\delta_\mathrm{\kappa=-(\ell+1)}\equiv\delta_\ell$) and compares them to the corresponding quantities obtained with $T=0$ K KS formula. The introduction of temperature and electronic correlations impacts mostly the low $\ell$ phase-shifts, i.e. in the present case, the $\ell=0$ and $\ell=1$ ones.
Figure \ref{figcumulxc} displays the cumulative integral
\begin{equation}\label{Ie_cumul}
    \int_0^\epsilon d\epsilon^\prime \left(-\dfrac{\partial f}{\partial \epsilon^\prime} \right)I(\epsilon^\prime) 
\end{equation}
and illustrates how the observed differences on the phase-shifts contribute to the electrical resistivity. One can notice that, despite the selection effect of the $-(\partial f/\partial\epsilon^\prime)$ term, not only the values of the phase-shifts in the vicinity of $\epsilon=\mu^*$ contribute, but also the low $\epsilon$ ones. 

\section{Impact of the ionic structure factor on Ziman's conductivity}\label{sec5}

\begin{figure}
     \centering
     \begin{subfigure}[b]{0.49\textwidth}
         \centering
         \includegraphics[width=\textwidth]{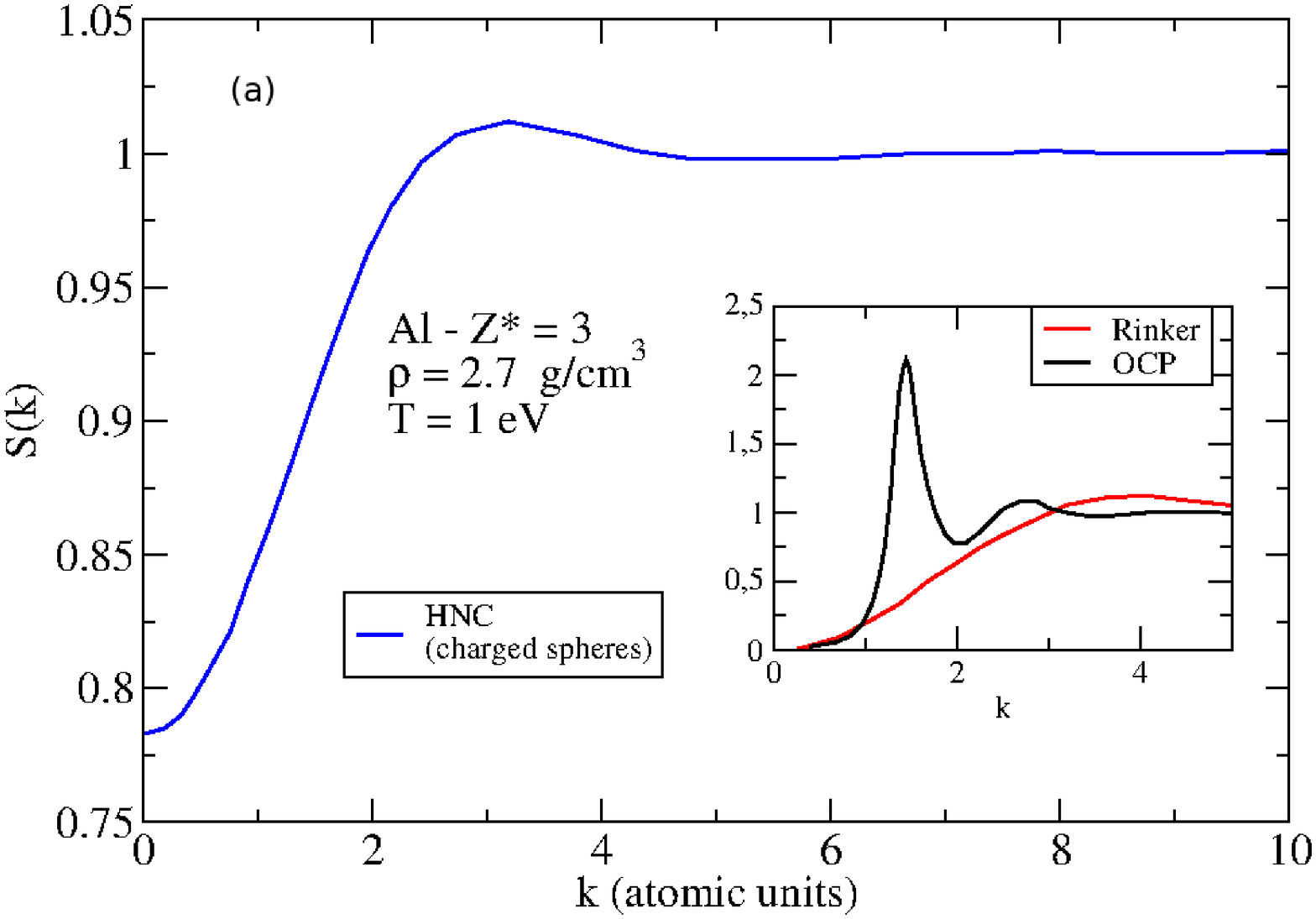}
     \end{subfigure}
     \hfill
     \begin{subfigure}[b]{0.49\textwidth}
         \centering
         \includegraphics[width=\textwidth]{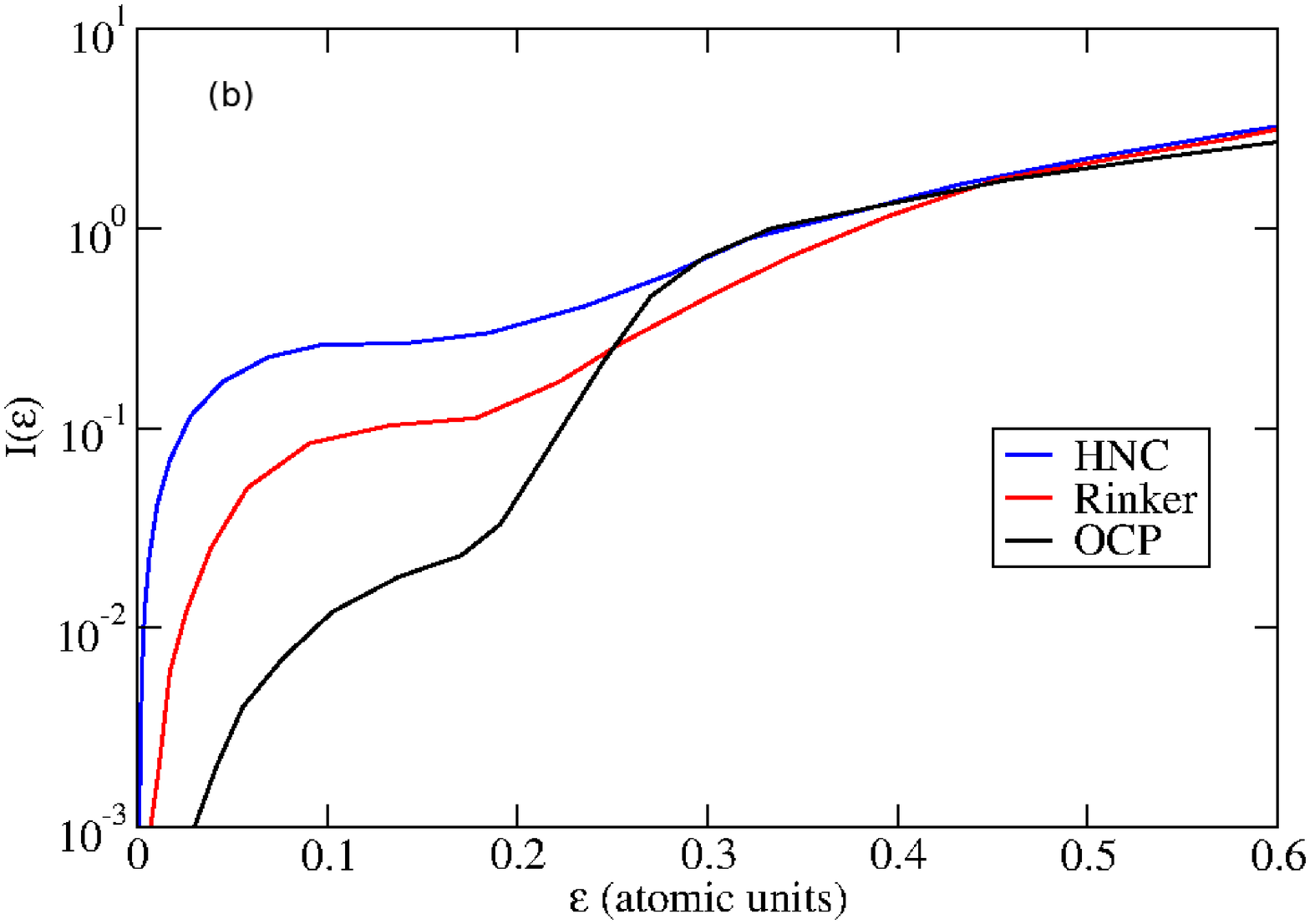}
     \end{subfigure}
     \hfill
\caption{(a): Solid density aluminum at $T=1$ eV: three models for the ionic structure factor. In blue: the HyperNetted-Chain solution for $Z^*=3$ charged spheres, used in our preceding work \cite{Wetta2020}. Inset: $S(k)$ from One-Component Plasma theory and from Rinker's formula (Equation (\ref{rinker})), both for coupling parameter $\Gamma=\beta Z^{*2}/R_\mathrm{WS}=82$, respectively presented in black and in red. (b): The result of the first integration step over the cosine $\chi$ of the deviation angle $\theta$.}
\label{figSk1}
\end{figure}
\begin{figure}
\centering
    \includegraphics[width=0.65\textwidth]{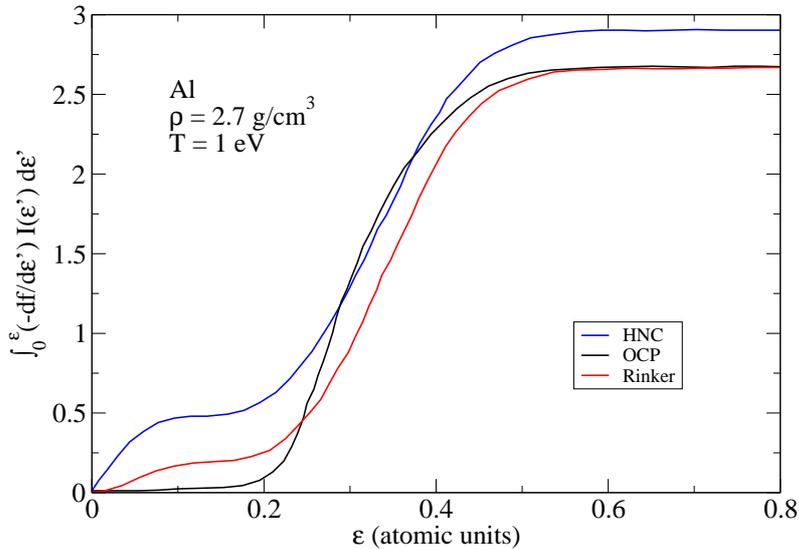}
    \caption{Cumulative integral $\int_0^\epsilon d\epsilon^\prime \left(-\frac{\partial f}{\partial\epsilon^\prime} \right)I(\epsilon^\prime)$: case of aluminum at $T=1$ eV. Impact of the ionic structure factor.
    \label{figcumulSk}}
\end{figure}

{\sc Paradisio} is based on the {\sc Inferno} model, which describes the central ion's environment by a jellium, i.e. an uniform electron gas, and an electrically compensating uniform distribution of charged ions. Therefore, in our preceding work \cite{Wetta2020}, the Ornstein-Zernike equations, together with the HNC closure relation, were solved for a system of spheres of charge $Z^*$, following Rogers's work \cite{rogers1980}. The calculations were initiated using the model direct correlation function of Held and Pignolet \cite{held1986}.

At low temperature ($T\lesssim$ 1 eV), the HNC solution tends to a system of loosely packed charged spheres, resulting in high $S(0)$ values, as shown in Figure \ref{figSk1}(a) for solid aluminum at $T=1$ eV. In such conditions, other approximations for $S(k)$ could be preferred. The inset presents two other possibilities: the black line corresponds to the One-Component Plasma (OCP) structure factor, as it is approximated by Desbiens {\it et al.} \cite{Desbiens2016}, and the red curve to Rinker's formula \cite{rinker1985}. The latter interpolates between the hard-sphere ionic structure factor and the Debye-H\"uckel one

\begin{equation}\label{rinker}
 S(k) = 1-\dfrac{3}{(kR_\mathrm{WS})^3} \left[\sin (kR_c)-(kR_c)\cos (kR_c) \right]-\dfrac{1-(R_c/R_\mathrm{WS})^3}{(kR_D)\left(1+R_c/R_D \right)}\dfrac{\sin (kR_c)+(kR_D)\sin (kR_c)}{1+(kR_D)^2}, 
\end{equation}
where $R_D=R_\mathrm{WS}/\sqrt{3\Gamma}$ denotes the Debye radius,  $\Gamma=\beta Z^{*2}/R_\mathrm{WS}$ being the plasma coupling parameter. Rinker proposed the following expression for the  core exclusion radius $R_c$
\begin{equation}
    \dfrac{R_\mathrm{WS}}{R_c}=\dfrac{1}{0.45^{1/3}}+\dfrac{2C}{\Gamma},
\end{equation}
where the adjustable parameter $C$ is chosen equal to 2, and 0.45 corresponds to the ideal packing fraction.

At first sight, the HNC structure factor seems odd, and one could think preferable to use the OCP or Rinker model presented in the inset. Actually, $S(k)$ impacts the resistivity $\eta$ through two successive integrations. The first one is performed over the cosine of the scattering angle $\theta$  to obtain $I(\epsilon)$, which is, in a second time integrated over $\epsilon$ (after multiplication by $-\partial f/\partial\epsilon$). According to Figure \ref{figSk1}(b), the HNC structure factor seems then, despite appearances, as plausible as the OCP and Rinker ones for the calculation of the resistivity.

Figure \ref{figcumulSk} underlines the importance of the low $k$ behavior of $S(k)$ and specifically its value $S(0)$ at the origin. The fact that $S(0)$ significantly differs from zero appears as an advantage that is hardly compensated by the fast raising of the OCP and Rinker structure factors. This suggests that, in order to obtain the high resistivities observed for plasmas and hot liquids, solving the Ornstein-Zernike equations should be preferred to the use of analytical models. 

We will develop further our analysis of the impact of $S(k)$ on the calculation of $\eta$ in the next section, through an analysis of a series of ultrafast heating experiments performed on aluminum.

\section{Sensitivity of the interpretation of two-temperature ultrafast experiments with respect to the ionic structure factor}\label{sec6}
\subsection{Two-temperature Ziman electrical resistivity}
%
Now we will employ our model developped in \cite{Wetta2020} to interpret experiments on ultrafast heating on aluminum. Milchberg {\it et al.} investigated the 1 eV $\lesssim T\lesssim$ 100 eV range by laser heating \cite{Milchberg1998}, while Sperling {\it et al.} measured aluminum's electrical conductivity heated to respectively 0.2 eV and 6 eV by x-rays, using the Linac Coherent Light Source (LCLS) facility \cite{Sperling2015}. In these ultrafast experiments, the sample is heated and probed on a timescale short enough to keep the ions at some temperature $T_i$ while electrons are heated up to $T_e\gg T_i$.

In the framework of the two-temperature Ziman formalism, the resistivity reads \cite{Dharmawardana1992,Petrov2018}
\begin{equation}
    \eta_\mathrm{2T}=-\dfrac{1}{3\pi}\dfrac{n_i(T_i)}{n_e(T_e)^2}\int_0^\infty\dfrac{\partial f}{\partial\epsilon}(\epsilon,\mu^*,T_e) I(\epsilon,T_i,T_e)d\epsilon.
\end{equation}
The electron density at $T_e$ and the ion density at $T_i$ being related by
\begin{equation}
    n_e(T_e)=Z^* n_i(T_i),
\end{equation}
one recovers the $1/(3\pi {Z^*}^2 n_i)$ factor. The integral $I(\epsilon,T_i,T_e)$ is given by
\begin{equation}
    I(\epsilon,T_i,T_e)=\int_0^{2k} q^3 S(q,T_i,T_e) \Sigma(q,T_e) dq.
\end{equation}
Relying on Petrov {\it et al.}'s observation that the ionic interactions weakly depend on the electron temperature in the case of aluminum\cite{Petrov2018}, we also assume that $S(q,T_i,T_e)\equiv S(q,T_i)$.
\subsection{Limits of our approach: the case of Sperling {\it et al.}'s x-ray heating experiments}
Sperling {\it et al.} measured surprisingly low electrical conductivities for aluminum at $T_e=$ 0.2 eV and $T_e=$ 6 eV, as compared to Milchberg {\it et al.}'s ones at $T_e\approx$ 0.8 eV.

\begin{figure}
    \centering
    \includegraphics[width=0.65\textwidth]{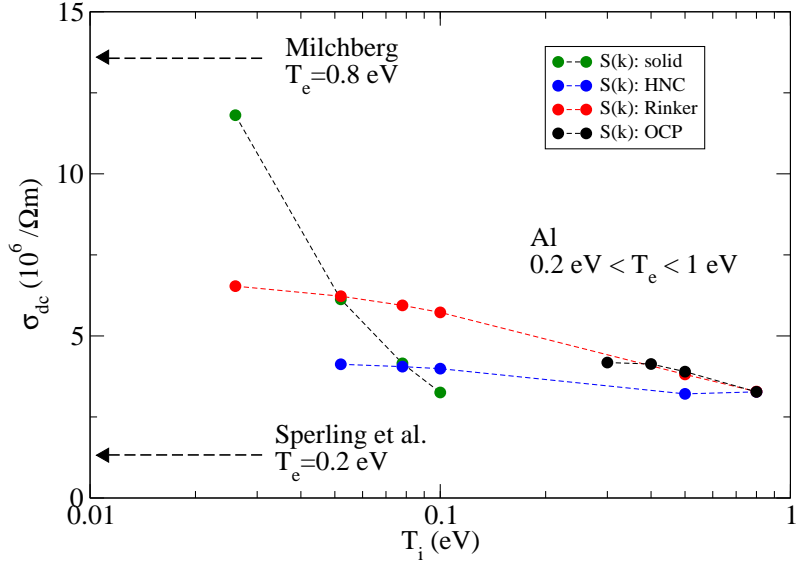}
    \caption{Two-temperature electrical conductivities: solid density aluminum at 0.2 eV $<T_i\leq T_e<$ 1 eV. Green circles: with solid-type $S(k)$. Blue, black and red circles: respectively with HyperNetted-Chain (HNC), One-Component Plasma (OCP), limited to $\Gamma\lesssim 200$, and Rinker's $S(k)$ (Equation (\ref{rinker})). The arrows indicate the values measured by Sperling {\it et al.}\cite{Sperling2015} and Milchberg {\it et al.}\cite{Milchberg1998}}
    \label{figmilchberg}
\end{figure}

Figure \ref{figmilchberg} presents our attempts to explain Sperling {\it et al.}'s $\sigma_\mathrm{dc}\approx 2\,10^6\,(\Omega\mathrm{m})^{-1}$ conductivity measured at $T_e=$ 0.2 eV, and Milchberg {\it et al.}'s $\sigma_\mathrm{dc}\approx 13\,10^6\,(\Omega\mathrm{m})^{-1}$ one at $T_e\approx$ 0.8 eV. At electronic temperatures varying in the range 0.2 eV $\lesssim T_e\lesssim$ 1 eV, we found that the two-temperature Ziman conductivity only varies with the ion temperature $T_i$ and the ionic structure factor $S(k)$. The green circles represent the conductivities obtained considering that the ions remain on the face-centered cubic (fcc) lattice sites. The corresponding structure factor was obtained according to the model (based on the Meisel-Cote-Debye approximation) developed in our preceding work \cite{Wetta2020} and detailed below. We also used experimental Debye temperatures measured for aluminum from ambient temperature up to melting \cite{Chipman1960}. Blue, black and red circles correspond to respectively HNC, OCP's and Rinker's approximations for liquid $S(k)$.

Sperling {\it et al.}'s experiment at $T_e=$ 0.2 eV can neither be interpreted considering a solid-like ionic $S(k)$, nor with that of a pure liquid. Actually, the experimental ionic structure factor, published by Witte {\it et al.} \cite{Witte2019}, presents a diffusive background characteristic of fluids, together with the Laue diffraction peaks associated to a fcc lattice. Modeling such an hybrid $S(k)$ is out of reach of our approach, unfit to account for complex interactions between ions on lattice sites, ``delocalised'' ones and electrons. Among the AA type approaches, only the NPA model adopted by Dharma-wardana {\it et al.} \cite{Dharmawardana2017} which solves consistently the electronic states and the ionic correlations, resulted in electrical conductivities consistent with the LCLS ones. Actually, their ion-ion structure factor is neither a solid one nor a pure liquid one, but a spherically averaged solid structure factor at $T_i=0.06$ eV resulting from the combination of a pseudo-potential built to account for phonon dispersion curves, to the Modified-HyperNetted-Chain equations.

Although we did not succeed in interpreting these results, our failure demonstrates the importance of the modeling of the ion-ion structure factor for that purpose.
\subsection{Case of Milchberg {\it et al.}'s laser heating experiments}
According to Figure \ref{figmilchberg}, the high electrical conductivity measured by Milchberg {\it et al.} at $T_e\approx$ 0.8 eV using fast laser heating technique can be explained with the aid of Ziman's two-temperature formalism considering that the ions remain on fcc lattice sites.

According to Rosenfeld and Stott \cite{Rosenfeld1990}, only the inelastic contributions (i.e. those involving at least one phonon in the electron-ion scattering) to the solid's structure factor contributes to the electrical resistivity. Therefore, in Ziman's expression, the total structure factor $S_\mathrm{tot}$ is replaced by the effective one
\begin{equation}
    S_\mathrm{eff}(k)=S_\mathrm{tot}(k)-S_\mathrm{el}(k),
\end{equation}
$S_\mathrm{el}(k)$ being the elastic contribution, i.e. in which no phonon is involved in the electron-ion scattering process. From Mangin {\it et al.}'s multi-phonon expansion\cite{Mangin1984}
\begin{equation}
    S_\mathrm{eff}(k)=\left[\sum_{n\ge 1} \frac{\left[2W(k)\right]^n}{n!}A_n(k)\right] \mathrm{e}^{-2W(k)}.
\end{equation}
Within the Debye model of the density of phonon states, $2W(k)$ reads 
\cite{kittel1963} 
\begin{equation}
    2W(k)=\dfrac{3\hbar^2 k^2}{m_i k_B}\dfrac{T^2}{\theta_D^3}\int_0^{\theta_D/T}\left[\dfrac{1}{e^t-1}+\dfrac{1}{2} \right]tdt, 
\end{equation}
$\theta_D$ denoting the Debye temperature, and $m_i$ the ion mass. Still applying the Debye model, the $A_n(k)$ functions are linked by a recursive relation\cite{Wetta2020}
\begin{equation}\label{recurence}
    A_n(k)= \dfrac{\int_0^{q_D}(2n_q+1)qdq\int_{-1}^1 A_{n-1}(k^2+\!\!q^2\!\!+2kq\chi)d\chi}{2\int_0^{q_D}(2n_q+1)qdq},
\end{equation}
$n_q$ denoting the phonon distribution function
\begin{equation}
    n_q=\dfrac{1}{\mathrm{e}^{\frac{\theta_D}{T}\frac{q}{q_D}}-1 },
\end{equation}
where $q_D$ reads
\begin{equation}
    q_D=\left(6\pi^2n_i\right)^{1/3}.
\end{equation}
The one-phonon function $A_1(k)$ reads\cite{Wetta2020}
\begin{equation}
    A_1(k)=\dfrac{\int_0^{q_D}(2n_q+1)qdq\sum\limits_G N(G)\int_{-1}^1 \delta(k^2+q^2+2kq\chi-G^2)\,d\chi}{4\pi\int_0^{q_D}(2n_q+1)qdq},
\end{equation}
where $N(G)$ is the number of reciprocal lattice vectors with the same length $G$. In its original form, the Meisel-Cote-Debye \cite{Meisel1977,Meisel1978} approximation assumes that
\begin{equation}
    A_n(k)=A_1(k)\ \ \text{for } n\geq 2,
\end{equation}
which leads to the following one-phonon structure factor
\begin{equation}
    S^{(1)}_\mathrm{eff}(k)=A_1(k)\left[1-\mathrm{e}^{-2W(k)}\right],
\end{equation}
where the superscript means that only the term $A_1(k)$ has been calculated. We extended the method by calculating up to $n$ functions $A_i(k)$, and assuming that
\begin{equation}
    A_m(k)=A_n(k)\ \ \text{for } m\geq (n+1),
\end{equation}
and subsequently the following recursive equation relating the multi-phonon structure factors
\begin{equation}
    S^{(n)}_\mathrm{eff}(k)-S^{(n-1)}_\mathrm{eff}(k)=\left[A_n(k)-A_{n-1}(k)\right]\left\{1-\mathrm{e}^{-2W(k)}\sum\limits_{i=0}^{n-1}\dfrac{\left[2W(k)\right]^i}{i!} \right\}.
\end{equation}
For solid density aluminum from ambient temperature up to melting, convergence is achieved with, at the most, $n=5$ functions $A_i(k)$.

\begin{figure}
    \centering
    \includegraphics[width=0.65\textwidth]{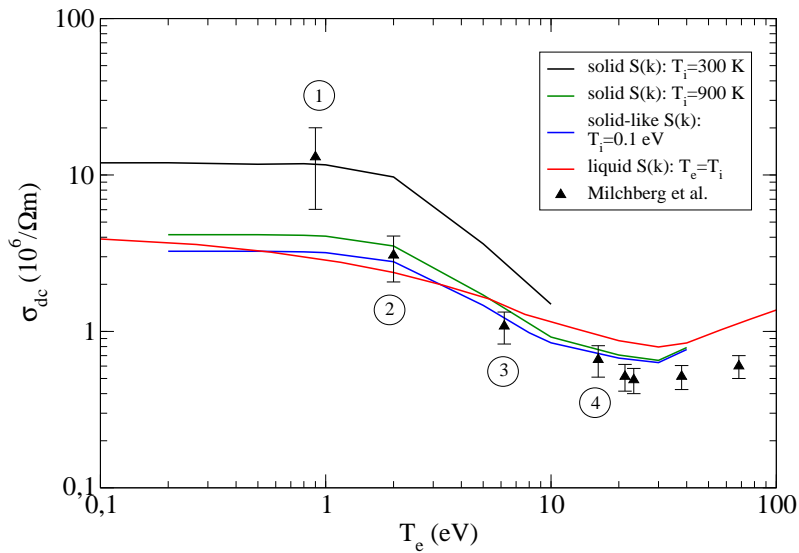}
    \caption{The figure compares Milchberg {\it et al.}'s laser heating experiments (triangles) to electrical conductivities calculated within the framework of the two-temperature Ziman formalism using solid ionic structure factors $S(k,T_i)$. Black curve:  with $T_i=$ 300 K, in green: $T_i=$ 900 K, and in blue: $T_i=0.1$ eV. The red curve recalls our $T_i=T_e$ calculations, assuming liquid structure factor.}
    \label{fig2T}
\end{figure}

Figure \ref{fig2T} presents the electrical conductivities obtained assuming ions on fcc lattices sites at $T_i=300$ K (black curve), 900 K (in green) and 0.1 eV (in blue), and compares them to Milchberg {\it et al.}'s experimental values. The calculations for $T_i=0.1$ eV (i.e. above melting temperature) correspond to an  attempt to simulate melted aluminum in which long-range fcc order still dominates over liquid one. The red curve corresponds to our $T_i=T_e$ (with liquid arrangement for the ions) calculations.

It appears that it is impossible to reproduce Milchberg's experimental values with a single $T_i$, however good agreement can be obtained when solid $S(k)$ at ion temperature $T_i$ progressively growing from 300 K to 0.1 eV is assumed. Specifically, Milchberg's point numbered 1 in the figure is consistent with $T_i=300$ K, point 2 with $T_i\gtrsim$ 900 K, where for points 3 and 4, $T_i$ must be greater than melting temperature. The experimental values of temperature $T_i$ are not available in paper of Milchberg {\it et al.}. We believe that the ion temperature $T_i$ may increase as $T_e$ grows. The results we present in Figure \ref{fig2T} confirm this.

Above $T_e\gtrsim$ 20 eV, all curves tend to the red one obtained considering thermal equilibrium $T_e=T_i$, and lay appreciably above the triangles representing the experiments. We interpret this discrepancy by the fact that the ions organize themselves progressively in a $T_i\ll T_e$ liquid order, as in Sperling {\it et al.}'s experiments. Like for the latter, our approach is unsuitable for the simulation of complex correlations between ions on fcc sites and delocalized ones.

\section{Conclusions}\label{sec7}

We studied the sensitivity of the resistivity $\eta$ to the different physical quantities required for its computation in the framework of Ziman's theory, such as the mean ionization $Z^*$ of the plasma, the exchange-correlation functional and the ion-ion structure factor $S(k)$.

The effect of $Z^*$ on $\eta$ is the most tricky to evaluate. Indeed, this parameter impacts $\eta$ by three ways. On one hand, it modifies the chemical potential and the ionic structure factor, and on the other hand, these two  effects are partially counterbalanced by the direct one on the $1/(3\pi {Z^*}^2 n_i)$ factor. There is probably no rule to guess the value of this compensation, making difficult to draw a clear recommendation for the definition of $Z^*$. However, it may be restrictive to consider that only the ideally free electrons (i.e evenly distributed outside the Wigner-Seitz sphere) contribute to the resistivity, and suggest to include all conduction electrons in $Z^*$.

The exchange-correlation effects has a clear impact on $\eta$ through the phase-shifts $\delta_\ell (k)$. We recommend to use the KSDT finite-temperature functional of Karasiev {\it et al.} (in its original form or with Groth {\it et al.}'s revised parameters), which allows the AA approach to calculate electrical resistivities in agreement with QMD ones for aluminum.

Finally, we studied the impact of the ion-ion structure factor on the resistivity. We observed that solving the Ornstein-Zernike equations, together with the HNC closure relation, provides the low $k$ behavior of $S(k)$ necessary to achieve agreement between Ziman resistivities and QMD ones. We showed the importance of the modeling of $S(k)$ in the study of ultrafast heating experiments. Thereby, agreement with Milchberg {\it et al.}'s laser heating experiments was only achieved with a solid state spherically averaged structure factor for the ions. We believe that Sperling {\it et al.}'s X-ray heating experiments could too be interpreted within the two-temperature Ziman formalism, with an ion structure factor including both liquid and solid aspects. 


\section*{Acknowledgments}
We are indebted to Stephanie Hansen for fruitful discussions about various aspects of the calculation of the dc conductivity within Ziman's formulation.

\end{document}